\documentclass[%
reprint,
amsmath,amssymb,
aps,
pra,
]{revtex4-2}
\usepackage{graphicx,graphics,color,epsfig}
\usepackage{bm}
\usepackage{amsmath}
\usepackage{mathrsfs}
\usepackage{amssymb}
\usepackage{subfigure}
\usepackage{amsfonts}
\usepackage{graphicx}
\usepackage{multirow,booktabs}
\usepackage{epstopdf}
\usepackage{diagbox}
\usepackage[colorlinks,
  anchorcolor=blue,  
  linkcolor=blue,       
  citecolor=blue]{hyperref}  
\usepackage{natbib}

\usepackage{dcolumn}
\usepackage{float}%
\usepackage{natbib}

\newcommand{\be}{\begin{eqnarray}}
\newcommand{\ee}{\end{eqnarray}}

\begin{document}
	
	\title{Quantum cluster kink and ring frustration}
	
	\author{Zhen-Yu Zheng}
	\affiliation{College of Physics, Sichuan University, 610064, Chengdu, People's Republic of China\\
		and Key Laboratory of High Energy Density Physics and Technology of Ministry of Education, Sichuan University, 610064,
		Chengdu, People's Republic of China}
	
	\author{Han-Chuan Kou}
	\affiliation{College of Physics, Sichuan University, 610064, Chengdu, People's Republic of China\\
		and Key Laboratory of High Energy Density Physics and Technology of Ministry of Education, Sichuan University, 610064,
		Chengdu, People's Republic of China}

	\author{Peng Li}
	\affiliation{College of Physics, Sichuan University, 610064, Chengdu, People's Republic of China\\
		and Key Laboratory of High Energy Density Physics and Technology of Ministry of Education, Sichuan University, 610064,
		Chengdu, People's Republic of China}
	\date{\today}
	
\begin{abstract}
  In this paper, we work on the pure and mixed cluster models with periodic boundary condition. The first purpose is to establish the concept of quantum cluster kink. We clarify that there are two types of cluster kinks since there are two types of ground states  depending on the choice of cluster length, of which the first type exhibits symmetry breaking order and the second one string order. Simple pictures are constructed according to the pure cluster model, which facilitates us to introduce the quantity, cluster kink number. As we demonstrate, cluster kinks deriving from different sources can coexist, compete with each other, and lead to quantum phase transition in a mixed cluster model. The second purpose is to elucidate that the effect of ring frustration can be realized in the cluster model with symmetry breaking order, instead of the one with string order. The reason lies in that ring frustration can induce a huge ground-state degeneracy or a special extended-kink phase with gapless excitations in the former case. And, although ring frustration does not change the phase transition point, it can produce unusual ground state in the extended-kink phase, whose special properties are uncovered by the nonlocal scaling factor in the correlation function and the doubled degeneracies of the eigenvalues of the entanglement spectrum.
\end{abstract}
	\maketitle

\makeatletter
\renewcommand{\numberline}[1]{%
	\settowidth\@tempdimb{#1\hspace{0.5em}}%
	\ifdim\@tempdima<\@tempdimb%
	\@tempdima=\@tempdimb%
	\fi%
	\hb@xt@\@tempdima{\@cftbsnum #1\@cftasnum\hfil}\@cftasnumb}
\makeatother
\newpage
	
\section{Introduction}

Kinks (or domain walls) play important roles in the field of condensed-matter physics \cite{nagaosabook,sachdevbook}. As the simplest type, Ising kinks occur in the classical Ising chain due to thermal fluctuations. But the scenario can be generalized to quantum spin chains, where Ising kinks can emerge as a consequence of quantum fluctuations. Moreover, quantum phase transition can be reflected by the behavior of increasing density of kinks in the ground state. The dynamics of kinks near a quantum critical point is featured by the well-known Kibble-Zurek mechanism \cite{zurek2005,Dziarmaga2005prl,zurek2007,campo2017,campo2019,najafi2020}. Various types of kinks can be defined according to the underlying models for practical problems ranging from magnetism \cite{vidal2020,Yin_2020,Yan_2021} to other research fields \cite{colemanbook}.

Recently, a family of cluster models have attracted lots of attention on the topics of symmetry protected topology and topological quantum computation \cite{pachos2004,smacchia2011,song2015,giampaolo2014,giampaolo2015,giampaolo2018,dutta2018,nie2017,
scaffidi2017,pollmann2017,zhang2018,ding2019,choo2018,azses2020,amico2010,wenbook}. In this work, we demonstrate that a general concept of  cluster kink can be established for these models. Then useful quantities, cluster kink number and its density, can be introduced to describe the quantum phase transition. More interesting, it turns out that there are two types of cluster kinks and the one with symmetry breaking order can provide the playground for exploring the effect of ring frustration \cite{dong2016,dong2018,li2019,zheng2019,Marco2020,gianpaolo2021,kou2021,Maric2021a,Maric2021b}.

The paper is organized as follows. In Sec. \ref{pure}, we establish the concept and picture of quantum cluster kink in an exact manner basing on the pure cluster Hamiltonian. Two types of kinks are introduced. The useful quantities, cluster kink number and its density, are defined. And the condition for realizing ring frustration is elucidated. In Sec. \ref{quantum}, we work on a mixed cluster model, which can possess a peculiar gapless extended-kink phase as an effect of ring frustration. The phase diagrams containing extended-kink phases are plotted. The density of kink number is shown to be a useful quantity for reflecting the competing orders in the ground states. Its second derivative exhibits a divergent peak at the critical point, whose scaling behavior has an interesting relation with the one of the second derivative of the ground-state energy density. The influences of ring frustration in the correlation function and entanglement spectrum are exemplified. At last, in Sec. \ref{conclusion}, we give brief summary and some discussions.

\section{Pure cluster model} \label{pure}

In this section, we focus on the pure $xzx$-$m$-cluster Hamiltonian with PBC ($N\gg m$),
\begin{align}
  H^{xzx}_m =J\sum_{j=1}^{N} \sigma_{j}^{x}\tau_{m,j}^{z}\sigma_{j+m}^{x}.  \label{Hcluster(m)}
\end{align}
where $\tau_{m,j}^{z}=\sigma_{j+1}^{z}\cdots\sigma_{j+m-1}^{z}$, the integer $m$ denotes the length of cluster interaction in each term. This Hamiltonian can be called ferro-cluster (FC) for $J<0$ and antiferro-cluster (AFC) for $J>0$ respectively.

It is noteworthy, in the classical Ising case ($m=1$), the system returns to the familiar form, $H^{xzx}_1\equiv H^{\text{Ising}}=J\sum_{j=1}^{N}\sigma_{j}^{x}\sigma_{j+1}^{x}$, which exhibits ferromagnetic (FM) order for $J<0$ and antiferromagnetic (AFM) order for $J>0$ correspondingly. Both orders break the $\mathbb{Z}_2$ symmetry of the Ising Hamiltonian.

For cases with $m>1$, the Hamiltonians are of quantum nature. It has been disclosed that the odevity of $m$ can influence the physical properties of the system \cite{wenbook}. We shall disclose the condition for realizing the effect of ring frustration by establishing the concept of quantum cluster kink.

\subsection{Quaternary Jordan-Wigner mapping}

The Hamiltonian in Eq. (\ref{Hcluster(m)}) with general $m$ can be easily handled by Jordan-Wigner transformation,
\begin{align}
	c_{j}^{\dagger}=\frac{1}{2}(\sigma_j^{x}+i\sigma_{j}^{y})\prod^{j-1}_{l=1}(-\sigma_{j}^{z}). \label{J-W}
\end{align}
But please notice the faithful \emph{quaternary Jordan-Wigner mapping} (QJWM) should be applied due to the presence of PBC. The complete mapping involves four Hamiltonians and can be pictorially expressed as \cite{zheng2019}.
\begin{align}
  \begin{array}{ccccc}
      &   & \boxed{H^{xzx}_m} &   &   \\
      & \nearrow &   & \nwarrow &   \\
                      \begin{array}{r}
                        H^{c}_{m} = \boxed{P_{z}^{-}H^{c}_{m}} \\
                        + \\
                        \boxed{P_{z}^{+}H^{c}_{m}}
                      \end{array}
     &   &   &   &
                      \begin{array}{l}
                        \boxed{P_{z}^{+}\tilde{H}^{c}_{m}} \\
                        + \\
                        \boxed{P_{z}^{-}\tilde{H}^{c}_{m}} = \tilde{H}^{c}_{m}.
                      \end{array} \\
      & \searrow &   & \swarrow &   \\
      &   & \boxed{\tilde{H}^{xzx}_m} &   &
  \end{array}
  \label{QJWM}
\end{align}
This mapping tells us the solution of $H^{xzx}_m$ can be decomposed into two fermion parity channels as,
\begin{equation}
  H^{xzx}_m = P_{z}^{-} H^{c}_{m} + P_{z}^{+} \tilde{H}^{c}_{m},
\end{equation}
where the fermion Hamiltonian with PBC ($c_{j+N}=c_{j}$) reads
\begin{align}
  H^{c}_{m}=(-1)^{m}J \sum_{j=1}^{N}(c_{j} - c_{j}^{\dagger})(c_{j+m} + c_{j+m}^{\dagger}),
\end{align}
while $\tilde{H}^{c}_{m}$ is a concomitant one with anti-PBC ($c_{j+N}=-c_{j}$). The parity projectors read
\begin{align}
  P_{z}^{\pm}=\frac{1}{2}(1\pm\mathscr{P}_{z}), \label{Pz}
\end{align}
with
\begin{align}
  &\mathscr{P}_{z} = \exp(i \pi M_z), \label{scrPz}\\
  &M_z = \sum_{j=1}^{N}\frac{1+\sigma_j^{z}}{2}=\sum_{j=1}^{N} c_{j}^{\dagger}c_{j}. \label{Mz}
\end{align}
It is noteworthy that the redundant degrees of freedom of $H^{c}_{m}$ and $\tilde{H}^{c}_{m}$ constitute the ones of $\tilde{H}^{xzx}_m$ as $\tilde{H}^{xzx}_m = P_{z}^{+} H^{c}_{m}+ P_{z}^{-} \tilde{H}^{c}_{m}$, which is also a pure $xzx$-$m$-cluster Hamiltonian but with anti-PBC.

\subsection{Picture of cluster kinks: symmetry breaking order and string order}

For general $m$, we can introduce a set of stabilizers for the ground state(s),
\begin{equation}
  S_{j}=\left\{\begin{array}{rl}
          \sigma_{j}^{x}\tau_{m,j}^{z}\sigma_{j+m}^{x} & (J<0,~\text{FC case}), \\
          -\sigma_{j}^{x}\tau_{m,j}^{z}\sigma_{j+m}^{x} & (J>0,~\text{AFC case}).
        \end{array}\right. \label{S_cluster}
\end{equation}
where the dependence of $S_{j}$ on $m$ is omitted for abbreviation. Usually, the ground state $|E_0\rangle$ is ordered and can be labelled by a set of uniform values $+1$,
\begin{align}
  S_{j}|E_0\rangle=+|E_0\rangle~~\forall j.
\end{align}
However, in the excited states, some values of stabilizers deviate from $+1$ to $-1$. When this occurs, we say \emph{cluster kinks} are created.

There is a distinct difference between the two cases, odd and even $m$ \cite{giampaolo2015,wenbook}. For $m\in$ odd, the doubly degenerate ground states exhibit $\mathbb{Z}_2$ symmetry breaking order, just like the classical Ising case ($m=1$). While for $m\in$ even, the unique ground state exhibits a string order without symmetry breaking.

Here, we draw the same conclusion naturally basing on the string of stabilizers ($r>m$),
\begin{align}
  \mathbb{S}_{j,r} =S_{j} S_{j+1}~\cdots~S_{j+r-m}. \label{S_string}
\end{align}
The pictures for two types of cluster kinks can be established subsequently. And we shall demonstrate that both types of cluster kinks can be faithfully labelled by the values of stabilizers.

\subsubsection{Type I cluster kink}

For $m\in$ odd, one can find that the string of stabilizers breaks into two disjoint local parts,
\begin{align}
  \mathbb{S}_{j,r} =[-\text{sgn}(J)]^{r}\mathcal{O}^{xy}_{j}~\mathcal{O}^{xy}_{j+r-m+1},
\end{align}
where each local part is a combined spin operator,
\begin{align}
  \mathcal{O}^{xy}_{j}=\sigma_{j}^{x}\sigma_{j+1}^{y}\cdots\sigma_{j+m-1}^{x}.
\end{align}
We also omit the dependence of $\mathbb{S}_{j,r}$ and $\mathcal{O}^{xy}_{j}$ on $m$ for abbreviation. Basing on the fact that the operator $\mathcal{O}^{xy}_{j}$ takes two possible eigenvalues, $\pm 1$, we define a correlation function for the ground state,
\begin{align}
  C^{\text{I}}(r)=[-\text{sgn}(J)]^{r}\langle \mathbb{S}_{j,r}\rangle =\langle \mathcal{O}^{xy}_{j}~\mathcal{O}^{xy}_{j+r-m+1}\rangle,\label{cfodd}
\end{align}
and the order parameter
\begin{align}
  \langle\mathcal{O}^{xy}_{j}\rangle = \sqrt{|C^{\text{I}}(r)|}, \label{OP}
\end{align}
to capture the $\mathbb{Z}_{2}$ symmetry breaking from the point of view of spin operators \cite{Kitaevbook}. Because we have $\langle \mathbb{S}_{j,r}\rangle=+1$ for the ordered ground states of the pure cluster Hamiltonian in Eq. (\ref{Hcluster(m)}), we get a simple result, $C^{\text{I}}(r)=[-\text{sgn}(J)]^{r}$.  In the Ising case, the definition of correlation function returns to the familiar form,
\begin{align}
  C^{\text{I}}(r) =\langle \sigma^{x}_{j}\sigma^{x}_{j+r}\rangle,
\end{align}
and the order parameter is usually marked by $\langle\sigma^{x}_{j}\rangle$.

\begin{figure}[t]
  \centering
  \includegraphics[width=7.5cm]{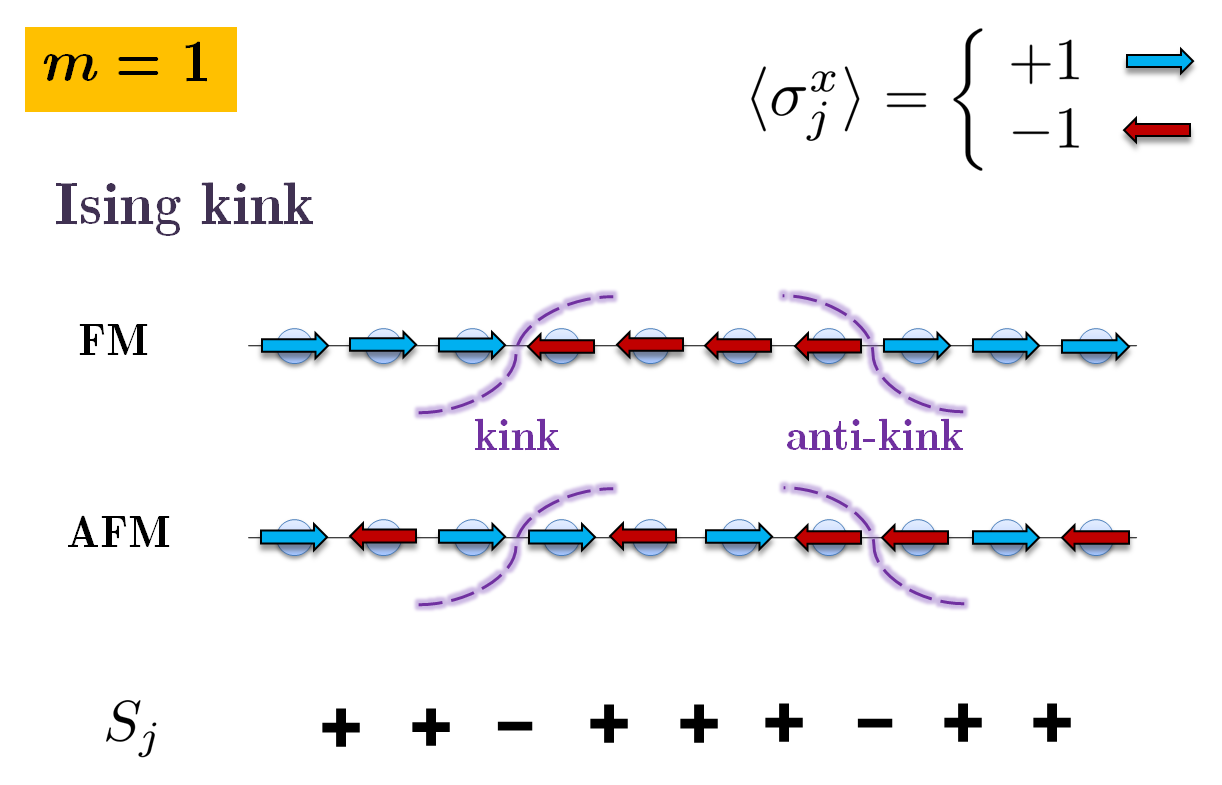}
  \caption{Ising kink. The picture basing on stabilizers $S_j$ is consistent with the one basing on the local order parameter $\langle\sigma_j^x\rangle$.}

  \label{Ising_kink} 
\end{figure}

\begin{figure}[t]
  \centering
  \includegraphics[width=7.5cm]{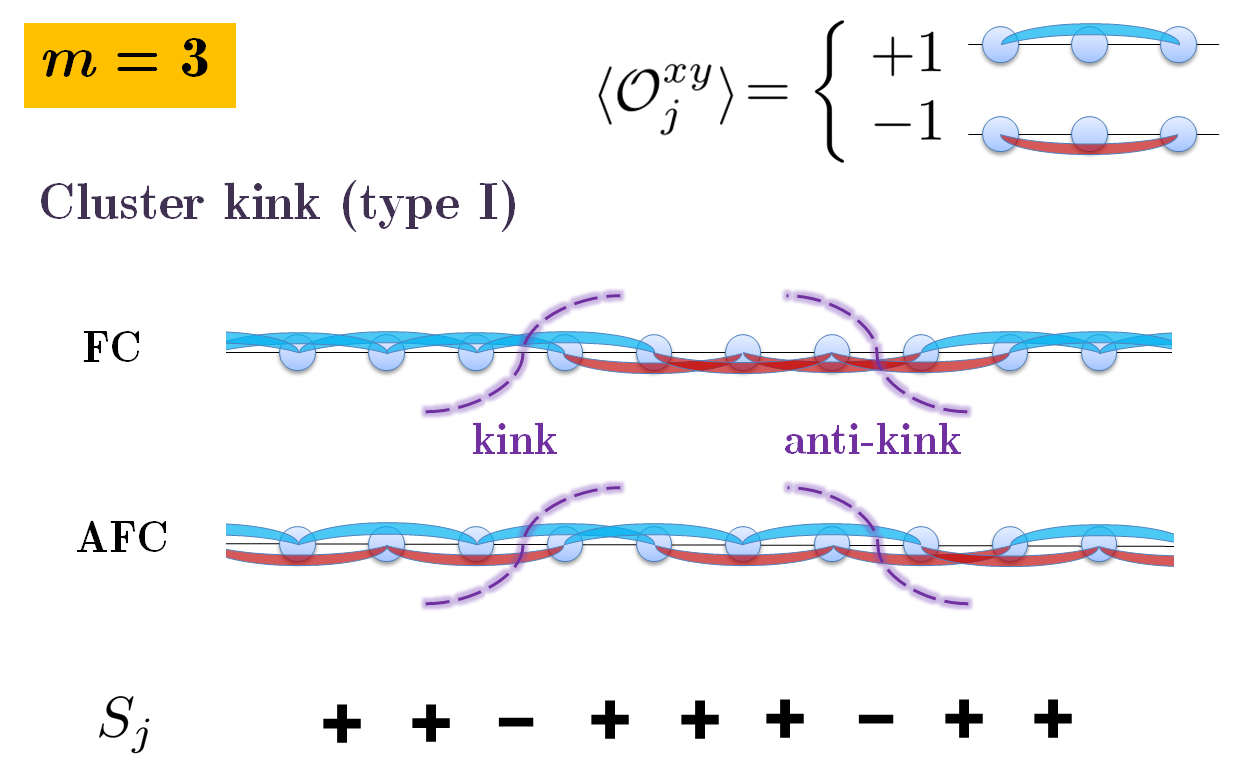}
  \caption{Type I cluster kink. Similar to the Ising case, the picture basing on stabilizers is consistent with the one basing on local order parameter $\langle\mathcal{O}^{xy}_{j}\rangle$. }

  \label{Cluster_kink_typeI} 
\end{figure}

Basing on the local order parameter, we can construct the picture for the cluster kink in the cases with $m\in$ odd. As the simplest example, let us see the picture of Ising kink as shown in Fig. \ref{Ising_kink}. We see that the picture basing on the stabilizers is consistent with the usual picture basing on order parameters. The picture for $m=3$ is illustrated in Fig. \ref{Cluster_kink_typeI}, which is a direct generalization of the Ising case in Fig. \ref{Ising_kink}. There are kink and anti-kink pairs in type I cluster kink.

\begin{figure}[t]
  \centering
  \includegraphics[width=7.5cm]{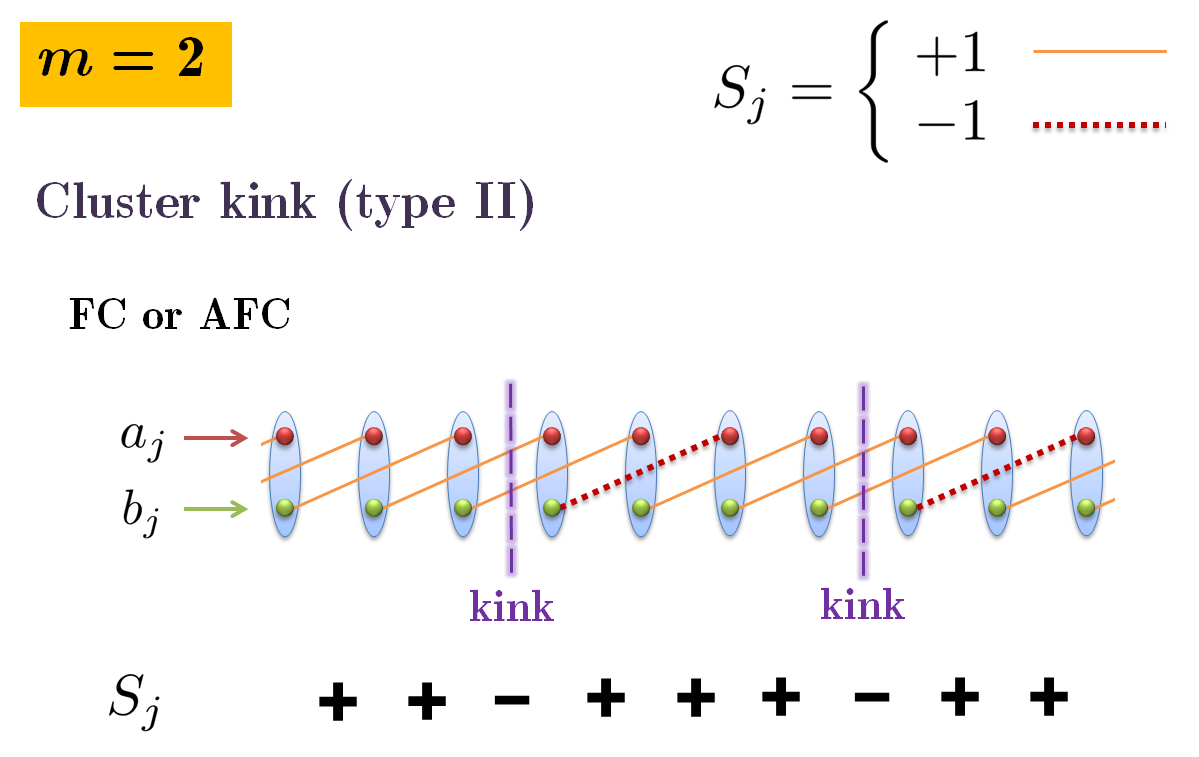}
  \caption{Type II cluster kink. The picture basing on stabilizers is consistent with the one basing on Majorana fermions. The relation between stabilizers and Majorana fermions can be found in Eq. (\ref{S_maj}). }

  \label{Cluster_kink_typeII} 
\end{figure}

\subsubsection{Type II cluster kink}

For $m\in$ even, the string can not be taken apart. For example, for $m=2$, we get
\begin{align}
  \mathbb{S}_{j,r} = [\text{sgn}(J)]^{r-3} \sigma^{x}_{j}\sigma^{y}_{j+1}(\sigma^{z}_{j+2}\cdots\sigma^{z}_{j+r-2})\sigma^{y}_{j+r-1}\sigma^{x}_{j+r}.
\end{align}
This means that one can not find a local order parameter from the point of view of spin operators and has to introduce a nonlocal string order that is represented by the whole string correlation function,
\begin{align}
  C^{\text{II}}(r)=[\text{sgn}(J)]^{r-3}\langle \mathbb{S}_{j,r}\rangle, \label{C_string}
\end{align}
which gives almost the same definition in previous studies \cite{giampaolo2015,ding2019}. Because we have $\langle \mathbb{S}_{j,r}\rangle=+1$, we get $C^{\text{II}}(r)=[\text{sgn}(J)]^{r-1}$ here. Interestingly, the new factor $[\text{sgn}(J)]^{r-1}$ implies that FC string order is staggered, while AFC string order is not.

Now we construct the picture for type II cluster kink. Since there is no local order parameter, we resort to the fermion language and transform the stabilizer to
\begin{align}
  S_{j}=\text{sgn}(J)(-1)^{m} i b_{j} a_{j+m}, \label{S_maj}
\end{align}
where
\begin{align}
  a_{j}=c_{j} + c_{j}^{\dagger},~~~b_{j} = i (c_{j}-c_{j}^{\dagger}),
\end{align}
are Majorana fermions. We can always pair the Majorana fermions coming from different lattice sites, $b_{j}$ and $a_{j+m}$, into a new fermion, $\tilde{c}_{j}$, so as to get \cite{kitaev2001}
\begin{align}
  S_{j}= 1 - 2\tilde{c}^{\dagger}_{j}\tilde{c}_{j}
       = \left\{\begin{array}{rl}
          +1~~~(\tilde{c}^{\dagger}_{j}\tilde{c}_{j}=+1), \\
          -1~~~(\tilde{c}^{\dagger}_{j}\tilde{c}_{j}=-1),
        \end{array}\right.
\end{align}
no matter in the FC or AFC cases. Thus the picture of the cluster kink basing on stabilizers is the same as that basing on Majorana fermions. The case for $m=2$ is illustrated in Fig. \ref{Cluster_kink_typeII}. One should notice that the uniqueness of the ground state prevents the occurring of spontaneous symmetry breaking.

\subsubsection{Calculation of the correlation functions}

We have defined two correlation functions in Eqs. (\ref{cfodd}) and (\ref{C_string}) for describing symmetry breaking order and string order respectively from the point of view of spin operators. However, their calculations are almost the same in the Majorana fermion language, because we have
\begin{align}
  C^{\text{I}}(r) = (-1)^{\frac{m-1}{2}}\langle b_{j}a_{j+m}...b_{j+r}a_{j+r+m}\rangle,
\end{align}
for the symmetry breaking order ($m\in$ odd), and
\begin{align}
  C^{\text{II}}(r)=(-1)^{\frac{m}{2}}\langle b_{j}a_{j+m}...b_{j+r}a_{j+r+m}\rangle,
\end{align}
for the string order ($m\in$ even). The same part, $\langle b_{j}a_{j+m}...b_{j+r}a_{j+r+m}\rangle$, can be decomposed by Wick's theorem and written in a Toeplitz determinant, which facilitates us to evaluate it by further analytical or numerical methods \cite{dong2016,li2019}.

\subsection{Cluster kink number}

Now we define the quantity for calculating number of kinks. For the pure cluster Hamiltonian in Eq. (\ref{Hcluster(m)}), the number of cluster kinks of a given excited state can be counted exactly since the state must be an eigenstate of the stabilizers. We define the number of cluster kinks as
\begin{align}
  \mathscr{N} = \frac{1}{2}\sum_{j=1}^{N}(1-\langle S_{j}\rangle), \label{N_K}
\end{align}
where $\langle S_{j} \rangle$ means the value of operator $S_{j}$ on a given state. Later, we shall demonstrate that the density of cluster kinks,
\begin{align}
  \rho = \frac{\mathscr{N}}{N}, \label{rho}
\end{align}
can depict the occurrence of quantum phase transitions. The saturate value of $\rho$ is $1/2$. The Ising case has been used in many previous works \cite{Dziarmaga2005prl}.

\subsection{The effect of ring frustration}

Two properties of the ground states of the Hamiltonian $H_{m}^{xzx}$ with $J>0$ (AFC case), the degeneracy $D$ and kink number $\mathscr{N}$ for sequences of $m$, are summarized in Table \ref{tableDN}. Basing on it, we shall elucidate the condition for the effect of ring frustration.

If one chooses $m\in$ even, the uniqueness of the ground state won't be changed no matter whether we set $N\in$ odd or $N\in$ even in the AFC case. Meanwhile the ground state possesses no kink that can be detected by the stabilizers. Nothing changes when one shifts the odevity of $N$. Thus there is no effect of ring frustration in these conditions. Another recent study came to the same conclusion \cite{Franchini}.

Whereas, the condition, $N\in$ odd and $m\in$ odd, is the right one for realizing ring frustration. First, the order parameter $\langle\mathcal{O}_{j}^{xy}\rangle$ takes alternative values, $+1$ and $-1$, just like the Ising case \cite{dong2016}. Second, the degeneracy of the ground states increase with the total number of lattice sites, $D=2N$. From the picture of type I cluster kink in Fig. \ref{Cluster_kink_typeI}, it is obvious to verify this fact pictorially. In a chain with perfect PBC, only $N-1$ stabilizers are independent and the last one is determined by them since we have
\begin{equation}
  S_{N}=-\prod_{j=1}^{N-1} S_{j}. \label{S_N}
\end{equation}
and the degeneracy reaches as large as $2N$ due to the translational symmetry of the Hamiltonian. All ground states are one-kink states because one stabilizer must take the value $-1$. Half of the ground states exhibit one kink and another half exhibit one anti-kink.

Now the system may fall into one of the $2N$ degenerate ground states and break both the $\mathbb{Z}_2$ symmetry and the translational symmetry. However, in the next section, we will show that quantum fluctuations can lift the degeneracy and lead to unique ground state with restored symmetries, meanwhile, a peculiar extended-kink phase emerges.

\begin{table}[!t]
  \centering
  \begin{tabular}{c|cc|c|cc||c|cc|c|cc}
		\hline
        \multicolumn{5}{l}{even $N$} && \multicolumn{5}{l}{odd $N$}
        \\
		\hline
		$\text{odd }m$&$D$&$\mathscr{N}$&$\text{even }m$&$D$&$\mathscr{N}$&
$\text{odd }m$&$D$&$\mathscr{N}$&$\text{even }m$&$D$&$\mathscr{N}$\\
        \hline
	    1&$2$&0 &2&1&0   &1&2$N$&1 &2&1&0\\
	    3&$2$&0 &4&1&0   &3&2$N$&1 &4&1&0\\
	    5&$2$&0 &6&1&0   &5&2$N$&1 &6&1&0\\
        $\vdots$&$\vdots$&$\vdots$&$\vdots$&$\vdots$&$\vdots$&$\vdots$&$\vdots$&$\vdots$&$\vdots$&$\vdots$&$\vdots$\\
		\hline
  \end{tabular}
  \caption{Degeneracy $D$ and kink number $\mathscr{N}$ of the ground states of the pure $m$-cluster Hamiltonian in Eq. (\ref{Hcluster(m)}) with $J>0$ (AFC case) for sequences of $m$. Please note the ground states whose degeneracy reads $D=2N$ due to the effect of ring frustration under the condition $N\in$ odd and $m\in$ odd.}
  \label{tableDN}
\end{table}

\subsection{Further remarks}

\subsubsection{Kramers-Wannier dual transformation}

To see the degeneracy of the ground state(s) in a constructive way, we can adopt the discussion basing on the Kramers-Wannier (KW) dual transformation alternatively \cite{pollmann2017}. The transformation should be designed appropriately according to the odevity of $m$ in order to fulfill the correct Pauli algebra. For $m\in$ even, the transformation should be defined as
\begin{align}
  Z_{j} = \sigma_{j-1}^{x}\tau_{m,j}^{z}\sigma_{j+m-1}^{x}, ~~X_{j} = \sigma_{j}^{x}, \label{KWTeven}
\end{align}
where the dependence of $Z_{j}$ on $m$ is omitted for abbreviation, so the pure cluster Hamiltonian can be rewritten in a dual form,
\begin{eqnarray}
  H_{m}^{xzx}=J\sum_{j=1}^{N}Z_{j}. \label{Hmeven}
\end{eqnarray}
While for $m\in$ odd, the transformation should be
\begin{align}
  Z_{j} = \sigma_{j}^{x}\tau_{m,j}^{z}\sigma_{j+m}^{x}~(j\neq N), ~~Z_{N}=\sigma^{x}_{N}\prod_{l=1}^{m-1} \sigma^{z}_{l}, \nonumber\\
  X_{j} = \prod_{k=1}^{j}\sigma^{x}_{k}\left(\prod_{l=k+1}^{k+m-2}\sigma^{z}_{l}\right)\sigma^{x}_{k+m-1},\label{KWTodd}
\end{align}
and we get
\begin{eqnarray}
  H_{m}^{xzx}=J\sum_{j=1}^{N-1}Z_{j}+J\prod_{j=1}^{N-1}Z_{j}.  \label{Hmodd}
\end{eqnarray}
Then it is easy to check the degeneracies of the ground states listed in Table \ref{tableDN} by Eqs. (\ref{Hmeven}) and (\ref{Hmodd}) following the discussion in Ref. \cite{pollmann2017}. Under the condition for ring frustration, $N\in$ odd and $m\in$ odd, we see clearly again the degeneracy of the ground states, $D=2N$.

\subsubsection{Applicability of cluster kink number}
For the pure cluster Hamiltonian Eq. (\ref{Hcluster(m)}), the cluster kinks mainly occur in the excited states, which would be a consequence of thermal fluctuations. However, the kinks can also emerge and pervade in the ground state as a consequence of quantum fluctuations aroused by noncommutative terms added to the Hamiltonian, such as the ones due to a transverse field \cite{sachdevbook}. We shall demonstrate that the definition of cluster kink number in Eq. (\ref{N_K}) is also valid in this situation. Moreover, cluster kinks deriving from different sources can coexist, compete with each other, and induce quantum phase transition. The relation between the kink density and quantum phase transition will be exemplified in Sec. III in detail.

\section{Mixed cluster model} \label{quantum}

In the last section, we pointed out that a general concept of quantum cluster kink can be introduced in the pure cluster model in an exact manner. We also pointed out the effect of ring frustration can be realized under the special condition, $N \in$ odd and $m \in$ odd, in the AFC case. In this section, we demonstrate that the concept quantum cluster kink can be used to identify quantum phase transitions. More important, we shall investigate the effect of ring frustration in the ground states by extensively analyzing the kink number, (string) correlation function, and entanglement spectrum. We will see that the occurrence of quantum phase transition can not be altered by ring frustration, but ring frustration leaves "fingerprint" in the (string) correlation function and entanglement spectrum as a non-local information.

\subsection{The Hamiltonian}

Now we consider the mixed cluster model \cite{ding2019},
\begin{align}
	H=H^{xzx}_{m}+H^{yzy}_{n}+H^{z},
\label{MCM}
\end{align}
where $H^{xzx}_{m}$ can be found in Eq. (\ref{Hcluster(m)}), $H^{yzy}_{n}=\lambda\sum_{j=1}^{N}\sigma_{j}^{y}\tau_{j,n}^{z}\sigma_{j+n}^{y}$ is also a pure cluster model, and $H^{z}=-g\sum_{j=1}^{N}\sigma_{j}^{z}$ is a term due to transverse field.

By the Jordan-Wigner transformation in Eq. (\ref{J-W}), the mixed cluster model is mapped to two fermion Hamiltonians, $H^{(\mp )}$, according to PBC ($c_{N+j} = c_{j}$) or anti-PBC ($c_{N+j} = - c_{j}$), that can be cast into the same expression,
\begin{align}
H^{(\mathscr{P}_{z})}&=\sum_{j=1}^{N-m} h_{j}^{J}-\mathscr{P}_{z}\sum_{j=1}^{m}b_{j}^{J}
-\sum_{j=1}^{N-n}h_{j}^{\lambda}\nonumber\\
&+\mathscr{P}_{z}\sum_{j=1}^{n}b_{j}^{\lambda}+\sum_{j=1}^{N}h_{j}^{z},
\end{align}
where
\begin{align}
&h_{j}^{J}=(-1)^{m}J(c_{j}-c_{j}^{\dagger})(c_{j+m} + c_{j+m}^{\dagger}),\nonumber\\
&h_{j}^{\lambda}=(-1)^{n}\lambda(c_{j} + c_{j}^{\dagger})(c_{j+n} - c_{j+n}^{\dagger}),\nonumber\\
&b_{j}^{J}=(-1)^{m}J(c_{N-m+j} - c_{N-m+j}^{\dagger})(c_{j} + c_{j}^{\dagger}),\nonumber\\
&b_{j}^{\lambda}=(-1)^{n}\lambda(c_{N-n+j} + c_{N-n+j}^{\dagger})(c_{j} - c_{j}^{\dagger}),\nonumber\\
&h_{j}^{z}= - g (2c_{j}^{\dagger}c_{j}-1).
\end{align}
The definition of $\mathscr{P}_{z}$ can be found in Eq. (\ref{scrPz}). By the Fourier transformation,
\begin{equation}
	c_{j}=\frac{1}{\sqrt{N}}\sum_{q}c_{q}\mathrm{e}^{-i q j},
\end{equation}
and the Bogoliubov transformation,
\begin{align}
	\eta_{q}=u_{q}c_{q}-\mathrm{i}v_{q}c_{-q}^{\dagger}~~(q\neq 0~\text{or}~\pi),
\end{align}
with
\begin{align}
    &u_{q}^2=\frac{1}{2}\left[1+\frac{\epsilon(q)}{\omega(q)}\right],v_{q}^2=\frac{1}{2}\left[1-\frac{\epsilon(q)}{\omega(q)}\right],u_{q}v_{q}=
	\frac{\Delta(q)}{\omega(q)},	\nonumber\\
    &\epsilon(q)=-(-1)^{m}J\cos{(mq)}+(-1)^{n}\lambda\cos{(nq)}-h,\nonumber\\
	&\Delta(q)=-(-1)^{m}J\sin{(mq)}-(-1)^{n}\lambda\sin{(nq)},\nonumber\\
    &\omega(q)=\sqrt{\epsilon(q)^2+\Delta(q)^2},\nonumber
\end{align}
the two free fermion Hamiltonians can be diagonalized in the momentum space as,
\begin{align}
	H^{(-)}&=\sum_{q\neq 0}\omega(q) (2\eta_{q}^{\dagger}\eta_{q}-1)\nonumber+C_{0}(2c_{0}^{\dagger}c_{0}-1),\\
	H^{(+)}&=\sum_{q\neq \pi}\omega(q) (2\eta_{q}^{\dagger}\eta_{q}-1)+C_{\pi}(2c_{\pi}^{\dagger}c_{\pi}-1),
\end{align}
where $C_{0}=-(-1)^{m}J-(-1)^{n}\lambda-h$, $C_{\pi}=-J-\lambda-h$. Notice that we have $q\in Q^{(-)}$ for $H^{(-)}$ and $q\in Q^{(+)}$ for $H^{(+)}$ with the definitions,
\begin{align}
	Q^{(-)}=\{-\frac{N-1}{N}\pi,..,-\frac{2}{N}\pi,0,\frac{2}{N}\pi,...,\frac{N-1}{N}\pi \},\\
    Q^{(+)}=\{-\frac{N-2}{N}\pi,..,-\frac{1}{N}\pi,\frac{1}{N}\pi,...,\frac{N-2}{N}\pi,\pi \}.
\end{align}

According to QJWM in Eq. (\ref{QJWM}), we can get the solution of $H$ by the projection,
\begin{equation}
	H = P_{z}^{-} H^{(-)} + P_{z}^{+} H^{(+)}.
\end{equation}
The BCS-like vacua of $H^{(-)}$ and $H^{(+)}$ are
\begin{align}
  &|\phi^{(-)}\rangle=\prod_{q\in Q^{(-)},0<q<\pi}(u_{q}+\mathrm{i}v_{q}c_{q}^{\dagger}c_{-q}^{\dagger})|0\rangle,\\
  &|\phi^{(+)}\rangle=\prod_{q\in Q^{(+)},0<q<\pi}(u_{q}+\mathrm{i}v_{q}c_{q}^{\dagger}c_{-q}^{\dagger})|0\rangle
\end{align}
respectively, where $|0\rangle$ is a full polarized state with spin down in $z$ direction. But notice that they are not the ground states of the problem. In fact, all energy states can be parsed after projection \cite{dong2016}.

\subsection{Extended-kink phases due to ring frustration}

\begin{figure}[t]
	\centering
	\subfigure[]{
		\includegraphics[width=7.5cm]{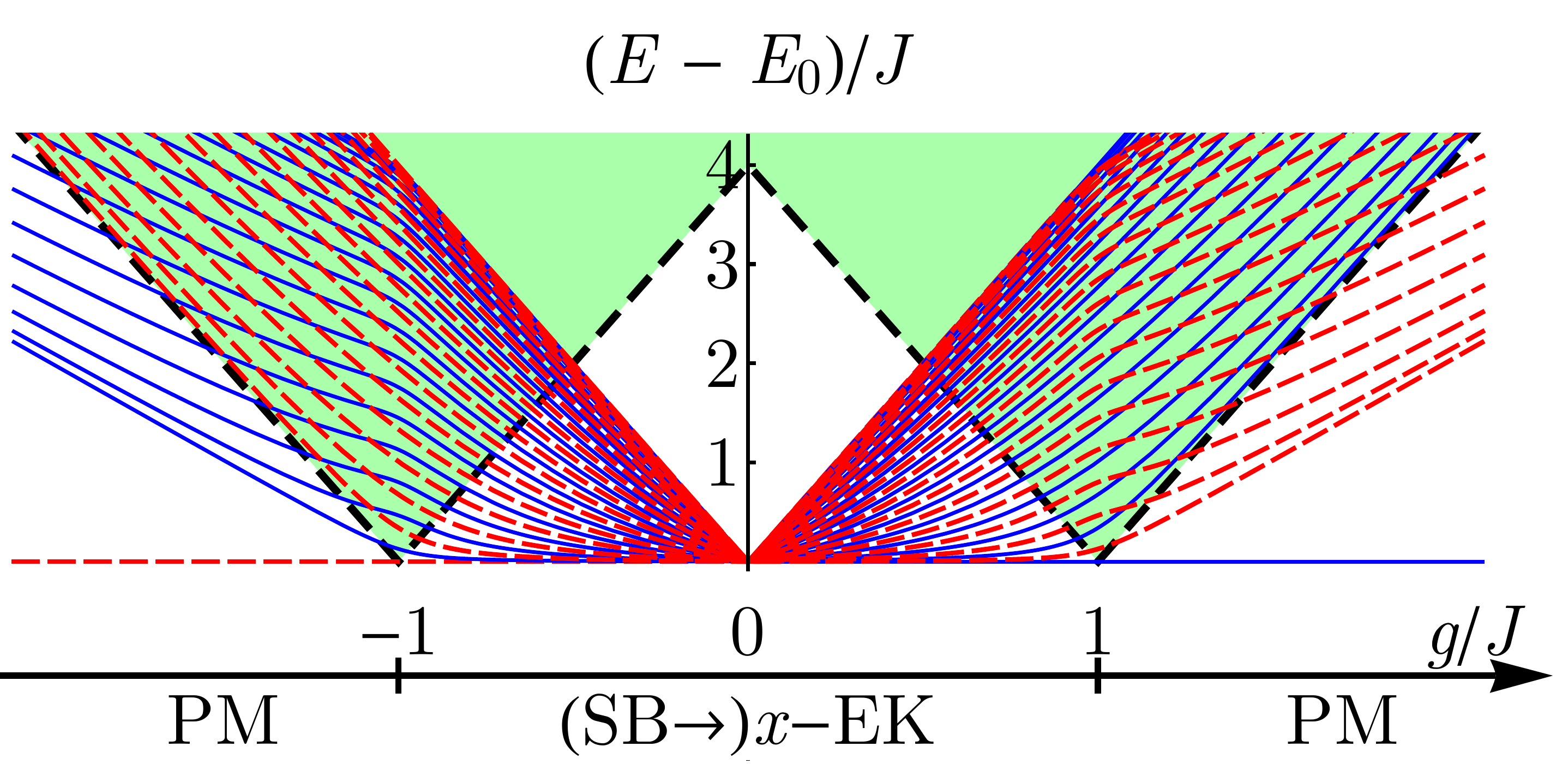}
	}
	\subfigure[]{
		\includegraphics[width=7.5cm]{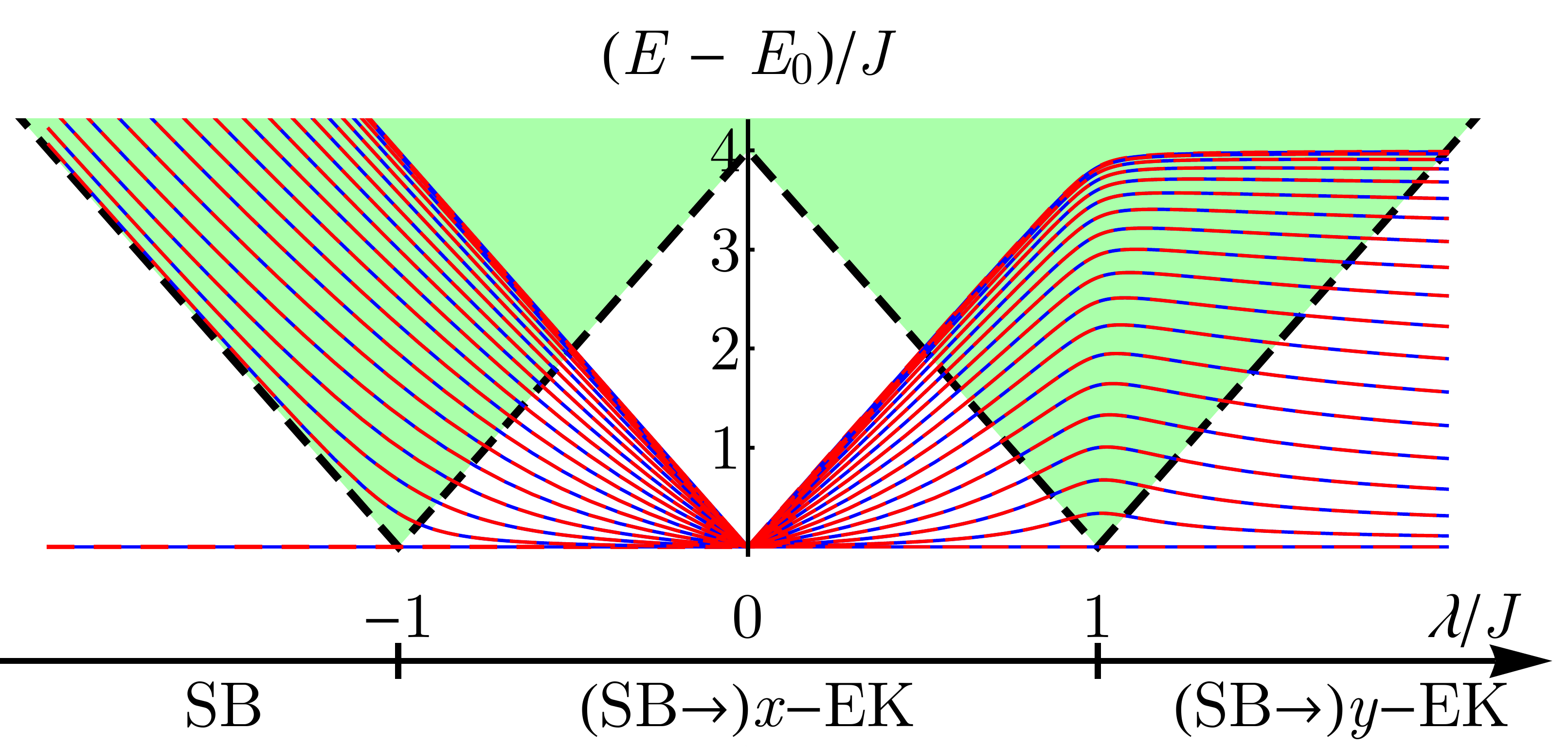}
	}
	\subfigure[]{
		\includegraphics[width=7.5cm]{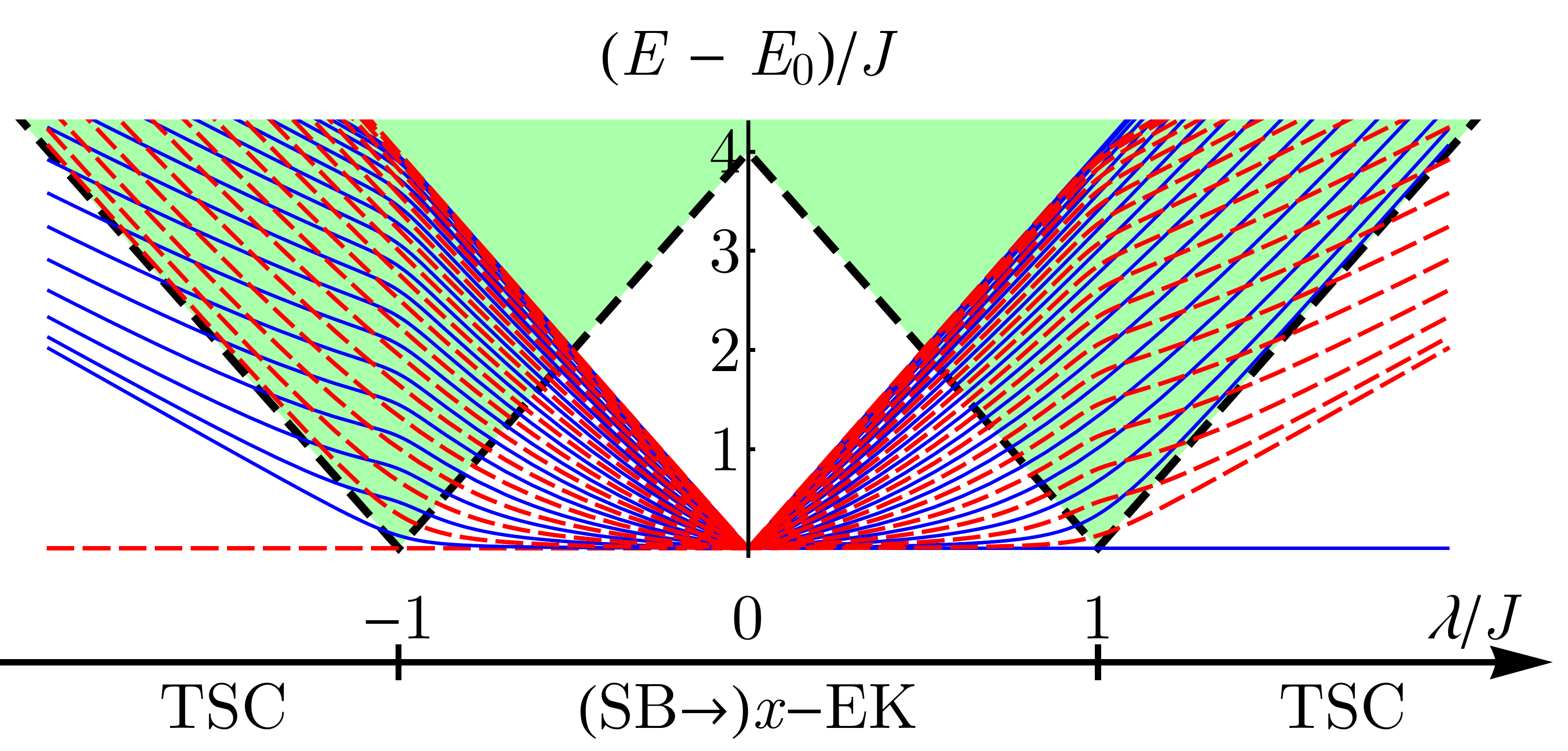}
	}
  \caption{Phase diagrams of the mixed cluster model with ring frustration ($N\in$ odd, $J>0$, and $m\in$ odd) in different cases: (a) $\lambda=0$ and $m =3$; (b) $g=0$, $m =3$, and $n =3$; (c) $g=0$, $m =3$, and $n= 2$. The gaplessness in the EK phases occurs in the thermodynamic limit, $N\rightarrow\infty$. Here, we choose $N = 37$ for demonstration. The lowest $2N$ energy levels are plotted along, which facilitates us to identify the gapless EK phases. The blue lines represent energy levels with odd parity, while the red dashed lines even parity. All other higher levels distribute in the shaded area above the black dashed line. The label "(SB$\rightarrow$)$x/y$-EK" means that the EK phase emerges by replacing a gapped SB phase (when $N\in$ even, i.e. without ring frustration) in the same parameter range.}
\label{PD}
\end{figure}

For the simple pure cluster model, the effect of ring frustration is reflected by the huge degeneracy of the ground states as large as $2N$ (Please see Table \ref{tableDN}). Now for the mixed cluster model, if the condition for ring frustration is still at presence (i.e. $N\in$ odd and $m\in$ odd), we demonstrate that gapless extended-kink (EK) phases emerge as the degeneracy is lifted by quantum fluctuations and the lowest gapless band is thus composed by the $2N$ low-lying energy levels \cite{dong2016}.

There are many cases for the EK phases to appear. Both $H_{m}^{xzx}$ and $H_{n}^{yzy}$ in the mixed cluster model can provide the source for EK phases, of which the former is favor of $x$-EK phase and the latter $y$-EK one. Let us exemplify them by focusing on three examples belonging to the general Hamiltonian in Eq. (\ref{MCM}): (a) $\lambda=0$ and $m=3$; (b) $g=0$, $m =3$, and $n =3$; (c) $g=0$, $m =3$, and $n=2$. The corresponding ground-state phase diagrams are illustrated in Fig. \ref{PD} one to one. In all examples, we have set $N\in$ odd and $J=1$ (energy unit) so that at least one EK phase appears.

In Fig. \ref{PD}(a), we observe two paramagnetic (PM) and one $x$-EK phases. In Fig. \ref{PD}(b), we get the symmetry breaking (SB), $x$-EK, and $y$-EK phases. Notice the SB phase exhibits doubly degenerate ground states and symmetry breaking order \cite{wenbook}. In Fig. \ref{PD}(c), we observe one $x$-EK and two topological superconducting (TSC) \cite{zheng2019} phases. In both TSC and PM phases, the ground state is unique. However, the ground-state degeneracy in the EK phases depends. In examples (a) and (c), the ground state of $x$-EK phase is unique. While, in example (b), the degeneracy is 2 for the $x$-EK phase with $-1<\lambda<0$ and 4 for the $x$-EK (and $y$-EK) phases with $\lambda>0$. Because the third example, (c), will be discussed extensively later, we write down it explicitly,
\begin{align}
  H&=H^{xzx}_{3}+H^{yzy}_{2} \nonumber\\
   &=\sum_{j=1}^{N}\sigma_{j}^{x}\sigma_{j+1}^{z}\sigma_{j+2}^{z}\sigma_{j+3}^{x}
      +\lambda\sum_{j=1}^{N}\sigma_{j}^{y}\sigma_{j+1}^{z}\sigma_{j+2}^{y},  \label{H32}
\end{align}
whose ground state reads
\begin{align}
  |E_0\rangle = \left\{
                \begin{array}{ll}
                  c_0^{\dagger}|\phi^{(-)}\rangle & (\lambda>0), \\
                  |\phi^{(+)}\rangle & (\lambda<0).
                \end{array}
                \right.
\end{align}

Overall, the ring frustration does not change the number of phases and the occurrence of critical point (the latter will be discussed in the next subsection). The EK phases are just replacements of the SB phases when the system's lattice shifts from $N\in$ even to $N\in$ odd. But the local order parameter disappears because the absence of symmetry breaking \cite{dong2016}.

\subsection{Kink number and quantum phase transition}

\subsubsection{Second derivative of kink number}

In the mixed cluster model, the two pure cluster terms, $H^{xzx}_{m}$ and $H^{yzy}_{n}$, apparently provide two competing types of cluster kinks. Correspondingly, we define two types of kink numbers, $\mathscr{N}^{xzx}_{m}$ and $\mathscr{N}^{yzy}_{n}$ (also two types of kink densities, $\rho^{xzx}_{m}$ and $\rho^{yzy}_{n}$), according to Eqs. (\ref{N_K}) and (\ref{rho}). Now we demonstrate that quantum phase transition can also be described from the point of view of competing cluster kinks, which gives the same conclusion of the critical point \cite{nie2017,ding2019}.

\begin{figure}[t]
	\begin{center}
	\subfigure[]{
		\includegraphics[width=7cm]{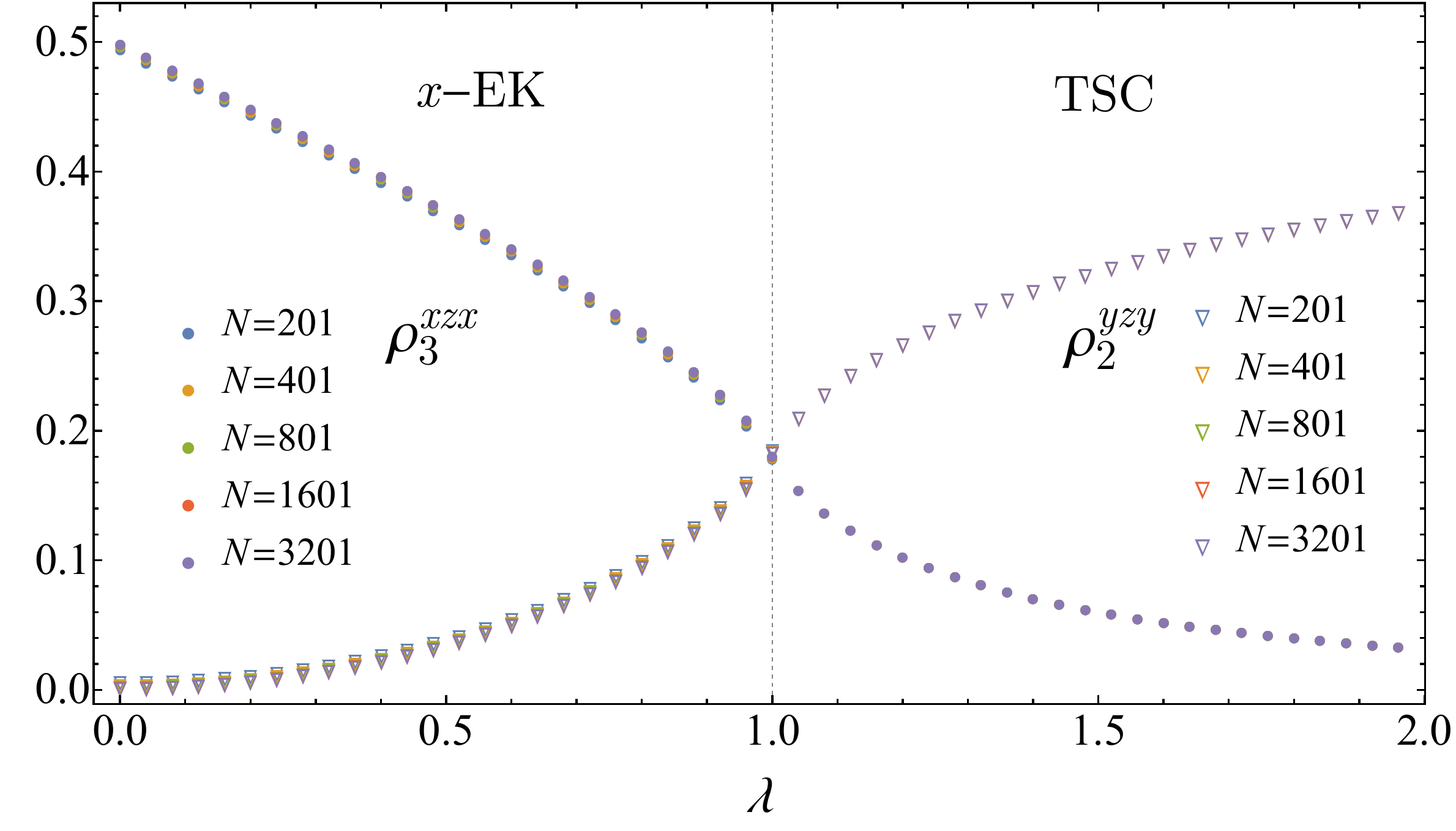}
	}
	\subfigure[]{
		\includegraphics[width=7cm]{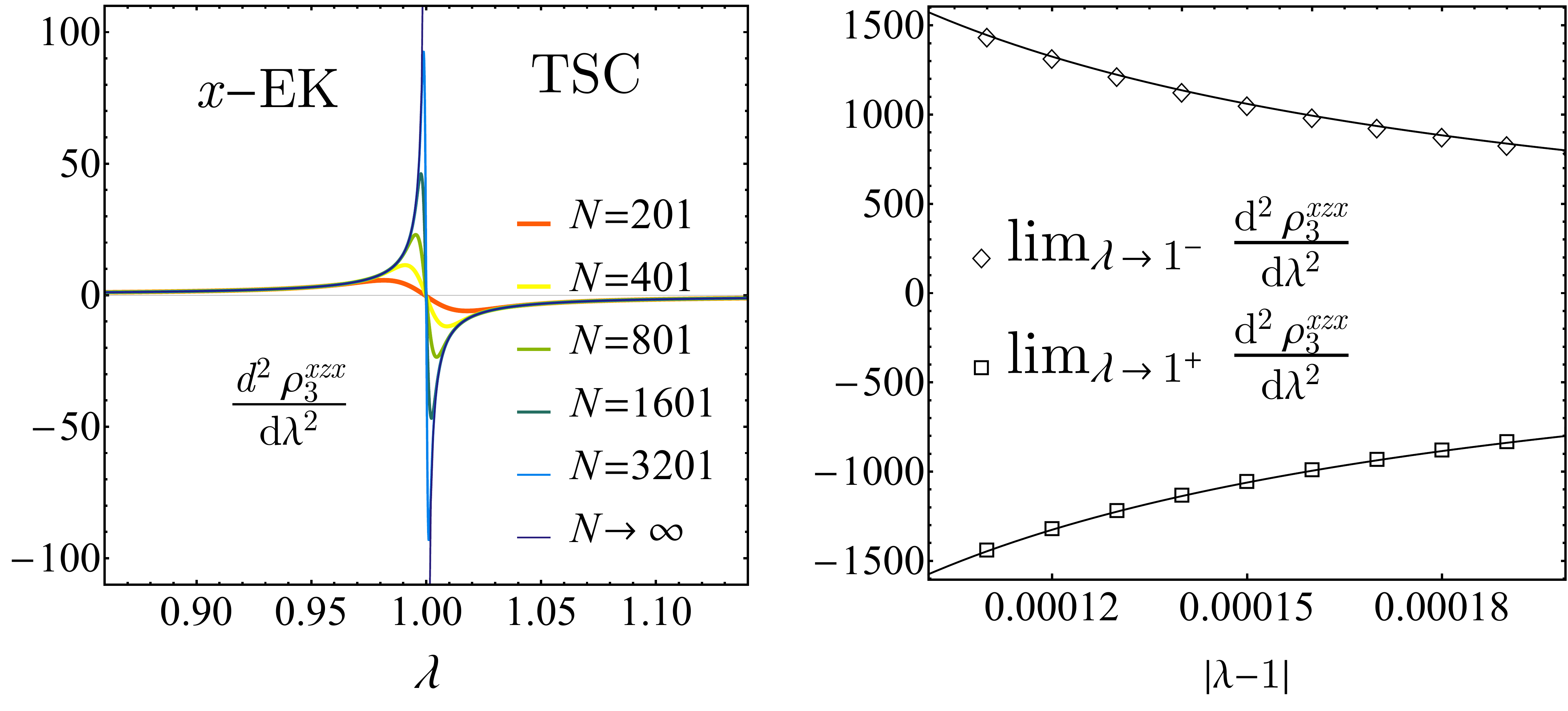}
	}
	\subfigure[]{
		\includegraphics[width=7cm]{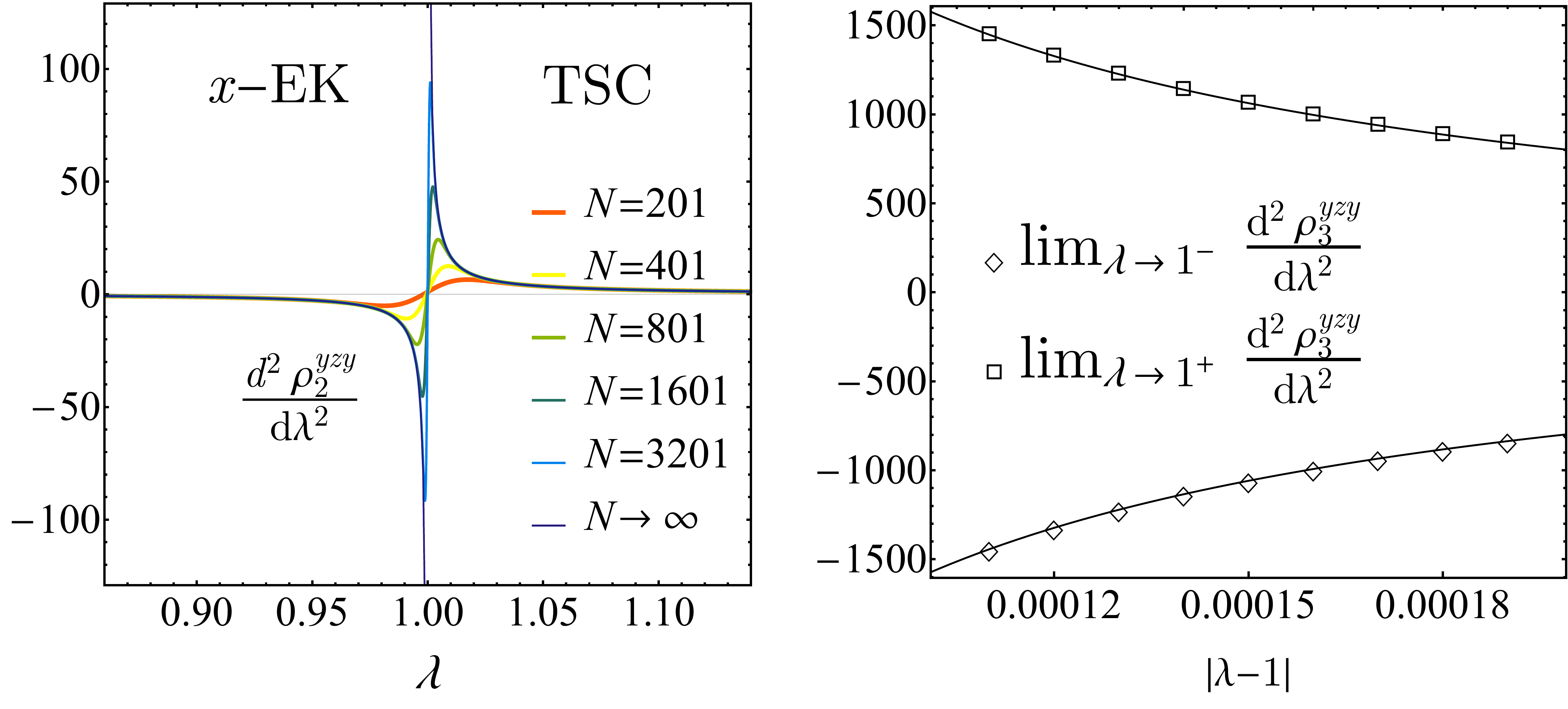}
	}
\end{center}
  \caption{(a) Kinks densities, $\rho^{xzx}_{3}$ and $\rho^{yzy}_{2}$, in the $x$-EK and TSC phases. (b) Left: The second derivative of kink density, $\frac{\mathrm{d}^{2}\rho^{yzy}_{2}}{\mathrm{d}\lambda^{2}}$; Right: Scaling analysis near the critical point. (c) Left: The second derivative of kink density, $\frac{\mathrm{d}^{2}\rho^{xzx}_{3}}{\mathrm{d}\lambda^{2}}$; Right: Scaling analysis near the critical point. Please see the fitting parameters in Eqs. (\ref{aa3}) and (\ref{aa2}). Here, we choose  the sequence of lattice sizes, $N=201, 401, 801, 1601, 3201$. }
	\label{Plot-Cluster-XY-kink-3}
\end{figure}

The idea can be borrowed from the transverse Ising model, which is the limiting case with $\lambda=0$ and $m=1$ in the mixed cluster model in Eq. (\ref{MCM}). In this limit, the kink density is in fact the energy density of the ground state, whose second derivative goes divergent at the critical point \cite{SuzukiBook}. Inspired by this precedent, we hope to use the second derivative of the kink densities, $\frac{\mathrm{d}^{2}\rho^{xzx}_m}{\mathrm{d}\lambda^{2}}$ (or $\frac{\mathrm{d}^{2}\rho^{yzy}_n}{\mathrm{d}\lambda^{2}}$), to characterize the phase transitions in the mixed cluster model. We expect the scaling behaviors,
\begin{align}
   &\frac{\mathrm{d}^{2}\rho^{xzx}_{m}}{\mathrm{d}\lambda^{2}}
   ~\left.(\text{or}~\frac{\mathrm{d}^{2}\rho^{yzy}_{n}}{\mathrm{d}\lambda^{2}} ) \right|_{\lambda\rightarrow\lambda_{c}\pm \delta}\nonumber\\
   &=\text{const}+a_{\pm}|\lambda-\lambda_{c}|^{-\kappa_{\pm}}+b_{\pm} \log|\lambda-\lambda_{c}|, \label{critical_behavior}
\end{align}
at the critical point $\lambda_{c}$, where $\delta=0^{+}$ is a infinitesimal and positive real number,  $\lambda\rightarrow\lambda_{c}\pm \delta$ means approaching the critical point from the left or right side, $\kappa_{+}$ and $\kappa_{-}$ are corresponding critical exponent. $a_{\pm}$ and $b_{\pm}$ are coefficients and can be obtained by data fitting. Usually, we needn't bother about the last logarithmic term if $\kappa_{\pm}\neq 0$.

To exemplify the idea, we study the Hamiltonian in Eq. (\ref{H32}). Notice that we set  $J=1$ and $N\in$ odd to impose ring frustration.
The phase diagram is depicted in Fig. \ref{PD}(c). The Hamiltonian has two competing orders, the symmetry breaking order and string order controlled by $H^{xzx}_{3}$ and $H^{yzy}_{2}$ respectively. When $-1<\lambda<1$, the system is in the $x$-EK phase, when $\lambda>1$, the system is in the TSC phase. There is a critical point at $\lambda_c = 1$. Let us observe the two types of kink densities,
\begin{align}
  &\rho^{xzx}_{3}=\frac{1}{2N}\sum_{j=1}^{N}(1+\langle\sigma_{j}^{x}\sigma_{j+1}^{z}\sigma_{j+2}^{z}\sigma_{j+3}^{x}\rangle),\\
  &\rho^{yzy}_{2}=\frac{1}{2N}\sum_{j=1}^{N}[1+\text{sgn}(\lambda)\langle\sigma_{j}^{y}\sigma_{j+1}^{z}\sigma_{j+2}^{y}\rangle],
\end{align}
and their second derivatives. Numerical results are illustrated in Fig. \ref{Plot-Cluster-XY-kink-3}. Basing on the data output from a sequence of lattice sizes, $N=201, 401, 801, 1601, 3201$, we observed the critical behaviors represented by the fitting parameters,
\begin{align}
  a_{-} = 0.158767,~a_{+} = -0.15879,~\kappa_{+}=\kappa_{-} = 1, \label{aa3}
\end{align}
for $\frac{\mathrm{d}^{2}\rho^{xzx}_3}{\mathrm{d}\lambda^{2}}$ and
\begin{align}
  a_{-} = -0.158744,~a_{+} = 0.158813,~\kappa_{+}=\kappa_{-} = 1, \label{aa2}
\end{align}
for $\frac{\mathrm{d}^{2}\rho^{yzy}_2}{\mathrm{d}\lambda^{2}}$ respectively. The values of $|a_{\pm}|$ are very close, so it seems safe to say that the divergent peaks are antisymmetric about the critical point (Please see Fig. \ref{Plot-Cluster-XY-kink-3} (b) and (c)).

\subsubsection{Relation between the kink numbers \\and the ground-state energy} \label{Relation}

There is a direct relation between the kink numbers and the ground state energy of the Hamiltonian in Eq. (\ref{H32}),
\begin{align}
  \varepsilon_{0}=2[\rho^{xzx}_{3}+\text{sgn}(\lambda)~\lambda\rho^{yzy}_{2}]
  -[1+\text{sgn}(\lambda)~\lambda], \label{epsilon0}
\end{align}
where $\epsilon_{0}=E_{0}/{N}$ is the ground-state energy density. It is well-known that the second derivative of the ground-state energy density, $\frac{\mathrm{d}^{2}\epsilon_{0}}{\mathrm{d}\lambda^{2}}$, can also capture the critical point, $\lambda_{c}=1$. However, by the scaling analysis on the data from a sequence of lattice sizes, $N=201, 401, 801, 1601, 3201$, we observed a symmetric logarithmic divergent behavior,
\begin{align}
	\frac{\mathrm{d}^{2}\varepsilon_{0}}{\mathrm{d}\lambda^{2}} \sim -0.31 \log|\lambda-1|,
\end{align}
which seems to contradict the relation in Eq. (\ref{epsilon0}) since  $\frac{\mathrm{d}^{2}\rho^{xzx}_3}{\mathrm{d}\lambda^{2}}$ and $\frac{\mathrm{d}^{2}\rho^{yzy}_2}{\mathrm{d}\lambda^{2}}$ exhibit power-law divergent peaks as disclosed in Eqs. (\ref{critical_behavior}), (\ref{aa3}), and (\ref{aa2}). In fact, this logarithmic divergent peak is a remain after the power-law divergent peaks in  $\frac{\mathrm{d}^{2}\rho^{xzx}_3}{\mathrm{d}\lambda^{2}}$ and $\frac{\mathrm{d}^{2}\rho^{yzy}_2}{\mathrm{d}\lambda^{2}}$ cancel with each other exactly. This reflects the interesting competition between the two types of cluster kinks at the critical point.

\subsubsection{Difference of kink number between \\with and without ring frustration}

However, if we consider the same Hamiltonian in Eq. (\ref{H32}) without ring frustration by setting $N\in$ even, the $x$-EK phase will be replaced by a SB phase with doubly degenerate ground states and gapped excitations. But the critical point still holds. Thus the kink density and its second derivative can not tell the difference between the $x$-EK and SB phases. To see how this happens, we investigate the kink number $\mathscr{N}^{xzx}_{3}$ for both systems with $N\in$ odd and $N\in$ even (i.e. the former exhibits ring frustration, the latter does not). We observe a robust behavior in the difference of kink numbers between them. For the Hamiltonian in Eq. (\ref{H32}), we can write down the result explicitly as
\begin{align}
  \Delta\mathscr{N}^{xzx}_{3} &= \mathscr{N}^{xzx}_{3}(N\in\text{odd}) - \mathscr{N}^{xzx}_{3}(N\in\text{even}) \nonumber\\
       &= \left\{
            \begin{array}{l}
              0~~~(\lambda<-1); \\
              1~~~(-1<\lambda<1); \\
              0~~~(\lambda>1).
            \end{array}
          \right.
\end{align}
This result can be verified by systems with various lattice sizes. In contrast, we can do the same calculation for the kink number $\mathscr{N}^{yzy}_{2}$, and get a trivial result,
\begin{align}
  \Delta\mathscr{N}^{yzy}_{2} &= \mathscr{N}^{yzy}_{2}(N\in\text{odd}) - \mathscr{N}^{yzy}_{2}(N\in\text{even}) \nonumber\\
       &= \left\{
            \begin{array}{l}
              0~~~(\lambda<-1); \\
              0~~~(-1<\lambda<1); \\
              0~~~(\lambda>1).
            \end{array}
          \right.
\end{align}
Thus the nontrivial value, $\Delta\mathscr{N}^{xzx}_{3}=1$ in the range $-1<\lambda<1$, is a good label for the effect of ring frustration.

\begin{widetext}
\begin{center}
\begin{table}[!h]
  \centering
  \begin{tabular}{c|c|c|c}
    \toprule
    \multicolumn{4}{l}{(a) $\lambda=0$ and $m\in$odd} \\
    \hline
      & PM ($g/J<-1$) & $x$-EK ($-1<g/J<1$) &  PM ($g/J>1$) \\
    \hline
    $C^{\text{I},xzx}_{m}(r)$ & 0 & $(-1)^{r+\frac{m}{2}-1}(1-\frac{h^{2}}{J^{2}})^{\frac{m}{4}}(1-2\alpha)$ & 0 \\
    \bottomrule
    \toprule
    \multicolumn{4}{l}{(b) $g=0$, $m\in$odd, and $n\in$odd} \\
    \hline
      & SB ($\lambda/J<-1$) & $x$-EK ($-1<\lambda/J<1$) &  $y$-EK ($\lambda/J>1$) \\
    \hline
    $C^{\text{I},xzx}_{m}(r)$ & 0 & $(-1)^{r+\frac{m}{2}-1}(1-\frac{\lambda^{2}}{J^{2}})^{\frac{m+n}{4}}(1-2\alpha)$ & 0 \\
    \hline
    $C^{\text{I},yzy}_{n}(r)$ & $(-1)^{\frac{n}{2}-1}(1-\frac{J^{2}}{\lambda^{2}})^{\frac{m+n}{4}}$ & 0 & $(-1)^{r+\frac{n}{2}-1}(1-\frac{J^{2}}{\lambda^{2}})^{\frac{m+n}{4}}(1-2\alpha)$ \\
    \bottomrule
    \toprule
    \multicolumn{4}{l}{(c) $g=0$, $m\in$odd, and $n\in$even} \\
    \hline
      & TSC ($\lambda/J<-1$) & $x$-EK ($-1<\lambda/J<1$) &  TSC ($\lambda/J>1$) \\
    \hline
    $C^{\text{I},xzx}_{m}(r)$ & 0 & $(-1)^{r+\frac{m}{2}-1}(1-\frac{\lambda^{2}}{J^{2}})^{\frac{m+n}{4}}(1-2\alpha)$ & 0 \\
    \hline
    $C^{\text{II},yzy}_{n}(r)$ & $(-1)^{\frac{n}{2}}(1-\frac{J^{2}}{\lambda^{2}})^{\frac{m+n}{4}}$ & 0 & $(-1)^{r+\frac{n}{2}}(1-\frac{J^{2}}{\lambda^{2}})^{\frac{m+n}{4}}$ \\
    \bottomrule
  \end{tabular}
  \caption{Correlation functions corresponding to the three typical phase diagrams in Fig. \ref{PD}. Please notice that there are nonlocal scaling factors, $(1-2\alpha)$ with $\alpha=r/N$, in the EK phases, which is a consequence of ring frustration.}
  \label{corr}
\end{table}
\end{center}
\end{widetext}

\subsection{Correlation function with nonlocal scaling factor}

It has been demonstrated that ring frustration manifests itself by leading to a nonlocal scaling factor in the correlation function \cite{dong2016,dong2018,li2019}. To capture it in the mixed cluster model, we work out the correlation functions, $C^{\text{I},xzx}_{m}(r)$ and $C^{\text{I/II},yzy}_{n}(r)$, according to the definitions in Eqs. (\ref{cfodd}) and (\ref{C_string}). To match the three typical cases of phase diagrams illustrated in Fig. \ref{PD}, we restore the subscript $m$, $n$, superscripts $xzx$, and $yzy$ to ascribe the sources of orders deriving from $H_{m}^{xzx}$ and $H_{n}^{yzy}$ respectively. The analytical results are listed in Table \ref{corr}, which clearly show that the EK phases are captured by the nonlocal scaling factor, $(1-2\alpha)$ with $\alpha = r/N$. The emergence of the nonlocal scaling factor marks the absence of symmetry breaking due to the effect of ring frustration \cite{kou2021}. Thus the nonzero correlation function does not mean any local order parameter in the EK phases. Instead, it reflects a nonlocal correlation without local order parameter just like the string correlation function.

\begin{figure}[t]
  \centering
  \includegraphics[width=8cm]{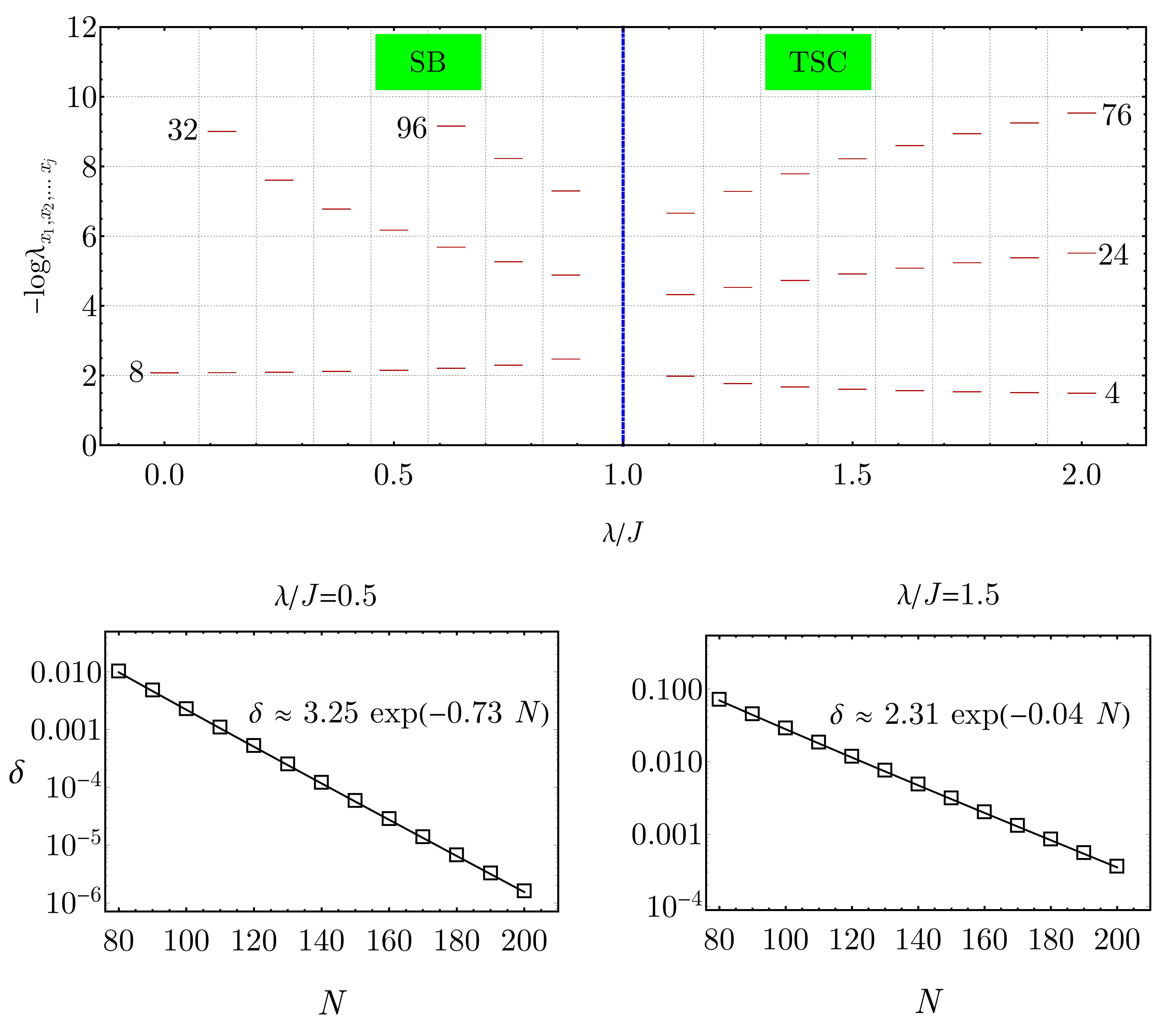}
  \caption{Upper plot: Entanglement spectrum of the Hamiltonian in Eq. (\ref{H32}) for $N=3000$ (no ring frustration). The degeneracies of three lowest eigenvalues are labelled. Lower left plot: Scaling analysis of the width $\delta$ of the lowest band of eigenvalues at $\lambda/J=0.5$. Lower right plot: Scaling analysis of the width $\delta$ of the lowest band of eigenvalues at $\lambda/J=1.5$. The scaling analyses indicate that the eigenvalues indeed tend to be degenerate in the thermodynamics. }
  \label{Plot_ES_even}
\end{figure}

\begin{figure}[t]
  \centering
  \includegraphics[width=8cm]{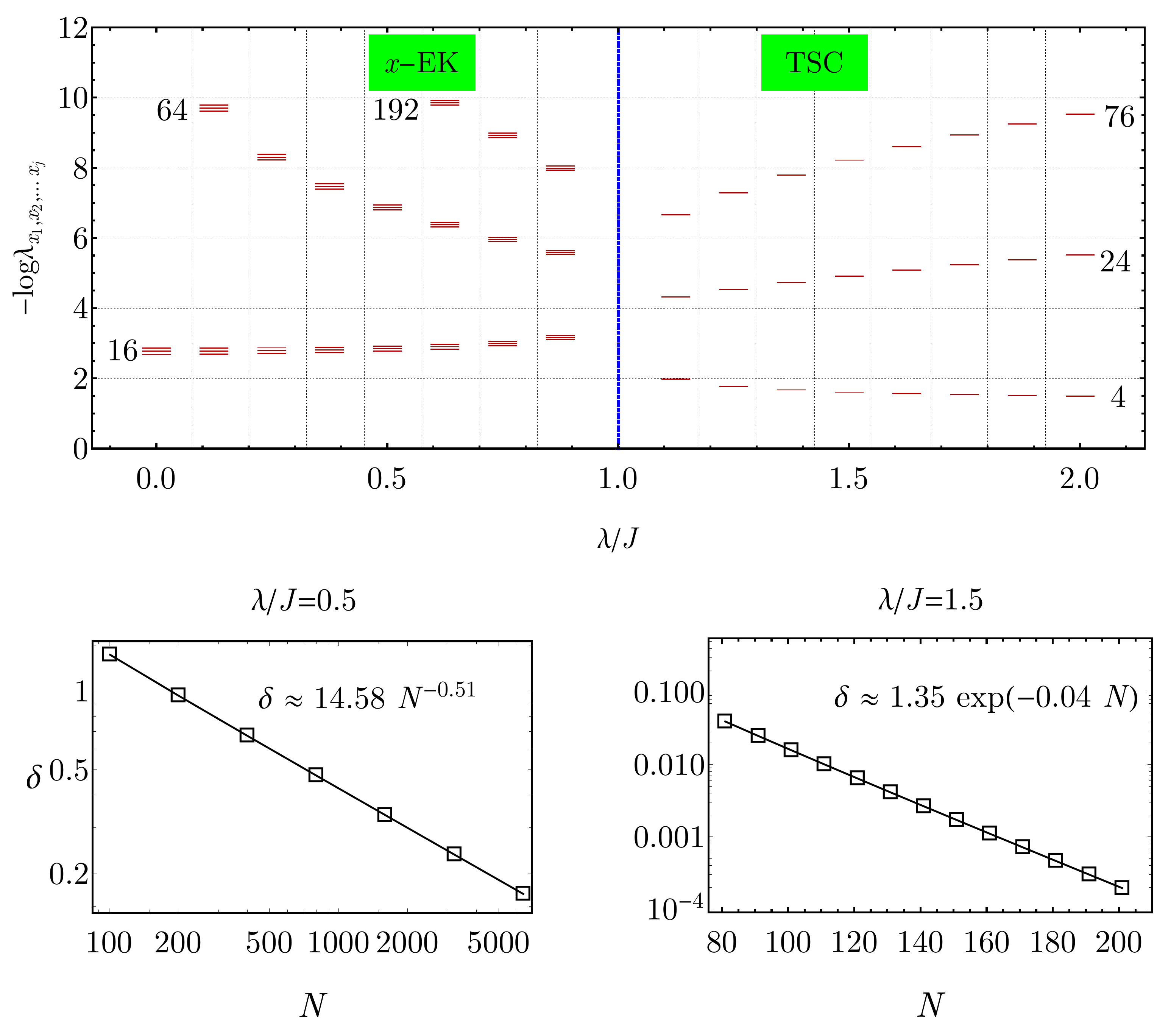}
  \caption{Upper plot: Entanglement spectrum of the Hamiltonian in Eq. (\ref{H32}) for $N=3001$ (with ring frustration). Lower left plot: Scaling analysis of the width $\delta$ of the lowest band of eigenvalues at $\lambda/J=0.5$. Lower right plot: Scaling analysis of the width $\delta$ of the lowest band of eigenvalues at $\lambda/J=1.5$. Notice that the degeneracies of the eigenvalues are doubled in the $x$-EK phase, where the scaling behavior is also quite different from that in the SB phase as shown in Fig. \ref{Plot_ES_even}. }
  \label{Plot_ES_odd}
\end{figure}

\subsection{Entanglement spectrum}

Entanglement spectrum \cite{li2008} is a powerful theoretical tool to analyze entanglement properties and topological
orders in exotic quantum states. The properties are reflected by the degeneracy of low-lying entanglement
spectrum \cite{fidkowski2010,pollmann2010}. Recently, the entanglement spectrum shows its importance in the investigation
of quantum phase transition \cite{calabrese2008,chiara2012,plat2020}, dynamical phase transition \cite{canovi2014,ueda2018,Jafari2021},
many-body localization phenomena \cite{serbyn2016}, as well as non-Hermitian systems \cite{ryu2020}.

The reduced density matrix of subsystem $l$ is defined by partial trace as
\begin{equation}
\rho_{l}=\mathrm{Tr}_{N-l}\rho,
\end{equation}
where $\rho$ is the full density matrix $|E_{0}\rangle\langle E_{0}|$.
And the entanglement spectrum of $\rho_{l}$ is the set of numbers $\lambda_{x_{1}x_{2}...x_{j}}$ which is defined as \cite{vidal2004}
\begin{equation}
	\lambda_{x_{1}x_{2}...x_{j}}=\prod_{j=1}^{l}\frac{1+(-1)^{x_{j}}v_{j}}{2},~~x_{j}=0,1 ~~\forall j.
\end{equation}
where $v_{j}$ is the imaginary part of the eigenvalues of the correlation matrix \cite{kitaev2003}.

Now we give the numerical results for the Hamiltonian in Eq. (\ref{H32}). To display the effect of ring frustration, we illustrate the results without ($N=3000$) and with ($N=3001$) ring frustration. When the system's size shifts from $N=3000$ to $N=3001$, SB phase is replaced by $x$-EK phase in the parameter range $0<\lambda/J<1$. We found that the degeneracies of eigenvalues in the $x$-EK phase is doubled compared with the ones in the SB phase (i.e. the numbers of eigenvalues of lowest two bands are 16 and 64 in the $x$-EK phase, while the ones are 8 and 32 in the SB phase). This phenomena can be taken as a fingerprint of ring frustration.

To confirm the degeneracies of the eigenvalues in the thermodynamical limit, we define $\delta$ as the width of the lowest band of gathered eigenvalues (for example, the lowest band is composed of 8 eigenvalues in the SB phase and 16 eigenvalues in the $x$-EK phase) and perform scaling analyses to see if it tends to become zero as the system's size increasing. The results are illustrated in Fig. \ref{Plot_ES_even} and \ref{Plot_ES_odd}. We conclude that the eigenvalues indeed tend to become degenerate in the thermodynamical limit. It is noteworthy that $\delta$ goes to zero algebraically in the $x$-EK phase, which is quite different from the exponential behaviors in other phases.

\section{Summary and discussion} \label{conclusion}

In summary, we have introduced the general concept of quantum cluster kink and pointed out the condition for realizing ring frustration in the pure and mixed cluster models. And as we have uncovered, there are two types of cluster kinks corresponding to two types of ground states, of which one has symmetry-breaking order and another nonlocal string order. In the former case, ring frustration can be realized, which induces the gapless EK phases in the mixed cluster models. Cluster kinks deriving from different sources can coexist, compete with each other, and lead to quantum phase transition. Although ring frustration does not change the phase transition point, it will bring a nonlocal scaling factor to the correlation function and double the degeneracy of the entanglement spectrum of the ground state in the EK phase.

From these peculiar conclusions, we see that ring frustration can provide us a brand new way to explore controllable interesting quantum extended-kink states with long-range correlation function but without symmetry breaking. As an odevity-induced phenomenon, the effect of ring frustration is reminiscent of the one in the well-known spin ladders \cite{Dagotto618}, which is rooted in the exotic parity structure of quantum states.

Nevertheless, there remain some uncovered aspects in such systems. First, for example, we can not cut the ring to maintain the effect of ring frustration, so the usual framework of bulk-edge correspondence that relates the bulk's nontrivial topology to the number of edge states \cite{Chiu} is not applicable here. Instead, one may resort to the bulk-defect correspondence \cite{kou2021}, but the general conclusion for the cluster model especially with high winding numbers \cite{song2015,nie2017,ding2019} is still lacking. Second, as a pending issue, the dynamics of cluster kinks is an interesting topic to go on with, since the general cluster kink exhibits a quantum nature that is different from the classical Ising kink \cite{zurek2005,Dziarmaga2005prl,torre2021odd}. Third, the definition of cluster kink relies on the specific terms in the Hamiltonian, so one would like to know whether it is applicable in other models without such terms like the ubiquitous Ising kinks.

\section*{ACKNOWLEDGMENTS}
The authors thanks F. Franchini and M. Giampaolo for useful discussion and information. This work is supported by NSFC under Grants No. 11074177.


%

\end{document}